\documentclass[aps,prl,twocolumn,showpacs,superscriptaddress,groupedaddress]{revtex4}  

\usepackage{graphicx}  
\usepackage{dcolumn}   
\usepackage{bm}        
\usepackage{amssymb}   
\usepackage[utf8]{inputenc}
\usepackage{color}
\usepackage[multidot]{grffile}
\usepackage{epstopdf}

\begin{document}

\widetext 
\leftline{Version 1 as of \today}
\leftline{Primary authors: Beno\^{\i}t Revenu and Vincent Marin}

\title{Radio emission from the air shower sudden death}
\centerline{author list dated 6 September 2012}
%

\affiliation{SUBATECH, Universit{\'e} de Nantes \'Ecole des Mines de Nantes IN2P3-CNRS, Nantes France.}

\author{V. Marin$^1$}
\author{B. Revenu$^1$}	

\date{\today}
\newcommand{\etal}{\MakeLowercase{\textit{et al. }}} 

\begin{abstract}
We present a new mechanism for air shower radio emission due to the sudden absorption of secondary particles when the shower front hits the ground. The electrons present in excess during the air shower development imply a net residual negative charge in the shower front. We show that for showers hitting the ground before the complete extinction of their electromagnetic component, the sudden vanishing of the net residual negative charge generates an electric field contribution in the kHz---MHz range. We characterize this radio contribution as a function of primary energy, arrival direction and antenna position, using the simulation code SELFAS2. We discuss the interest of this new predicted signal on detection and analysis of ultra-high energy cosmic-rays and we argue that the region in the shower of maximum emission of the electric field should not coincide with the region of maximum development.
\end{abstract}

\maketitle

\section{I. Introduction}
The electromagnetic particles of air showers reaching the ground are detected by ground particles detectors such as Water Cerenkov Detector (used at the Pierre Auger Observatory for instance~\cite{Allekotte:2007sf}) or plastic scintillators (used in the Telescope Array experiment for instance~\cite{AbuZayyad201287}). The measurement of their ground density allows an estimation of the core position and the primary energy. These particles should be also detectable through the electric field emitted consecutively to their sudden absorption in the ground. We propose in this paper a complete characterization of this signal using simulations.
We use the code SELFAS2~\cite{Marin2012733}, where each secondary electron and positron of the shower front is considered as a moving source. The total electric field emitted by the complete shower at any observation point $\boldsymbol{x}$, is obtained after superposition of all individual contributions. Considering only the radiative part of the field (we neglect the first term of Eq.~(17) in \cite{Marin2012733}), the total field detected by an observer located at $\boldsymbol{x}$ and at time $t$ is given by:
\begin{equation}
\boldsymbol{E}_{tot}(\boldsymbol{x},t)=\frac{1}{4\pi\varepsilon_0 c}\frac{\partial}{\partial t}\sum _{i=1} ^{N_t}q_i(t_{\text{ret}})\left[\frac{\boldsymbol{n}_i-\boldsymbol{\beta_i}}{R_i(1-\eta\boldsymbol{\beta}_i.\boldsymbol{n}_i)}\right]_{\text{ret}}
\label{SumField}
\end{equation}
where $\eta$ is the air refractive index, $\boldsymbol{n}_i$ and $R_i$ are the line of sight and the distance between the observer and the particle $i$, $\boldsymbol{\beta}_i$ the velocity of this particle and $q_i$ its electric charge. The summation is done over the total number $N_t$ of particles that emitted an electric field detected by the observer at time~$t$.
All these quantities are evaluated at the retarded time $t_{\text{ret}}$, related to the observer's time $t$ through:
\begin{equation}
t=t_{\text{ret}}+\frac{\eta \,R_i(t_{\text{ret}})}{c}\cdot
\label{RetTime}
\end{equation}
The Earth's magnetic field induces a systematic opposite drift of electrons and positrons during the shower development; this generates a residual current $\boldsymbol{j}_{\perp}=q\,c\,\boldsymbol{\beta}^{e^\pm}_{\perp}$ perpendicular to the shower axis. The variation of this current with the number of particles creates the main electric field contribution, in the MHz range. In addition, the charge excess variation due to positrons annihilation and knock-on electrons during the air shower development (Askaryan effect in air~\cite{ask1962}) gives a secondary contribution to the total electric field, also in the MHz range (see \cite{Marin2012733} for a complete description). Experimental evidence for this specific contribution can be found in~\cite{Schoorlemmer2012S134,marinicrc2011}. Finally, the air refractive index alters the radio signal for observers located close to the shower axis (less than 200~m for a vertical shower, see~\cite{PhysRevLett.107.061101}). The resulting electric field generated during the shower development in the atmosphere creates the principal pulse (PP), as already observed since years by several experiments such as CODALEMA~\cite{Ardouin:2005xm,coda2009},  LOPES~\cite{lopes1,Huege2012S72} or AERA~\cite{Abreu:2011ph} for instance.
Since the last decade, the convergence of various experimental results and various theoretical approaches shows that the mechanism of the radio signal emitted by the shower during its development in the atmosphere is quite well understood.

We suggest a new mechanism generating a clear radio signal due to the absorption of the shower front when reaching the ground. In this paper, we first describe the mechanisms at the source of this electric field. Then, we present the characteristics of this signal using simulated showers with SELFAS2, in both time and frequency domains. The dependence with the arrival direction, energy of the primary cosmic ray and the relative position of the observer with respect to the shower core is discussed. Finally, we discuss the importance of this new predicted signal in the field of ultra-high energy cosmic rays.

\section{II. Sudden death radio emission}

\begin{figure}[!ht]
\begin{center}
\includegraphics[scale=0.23]{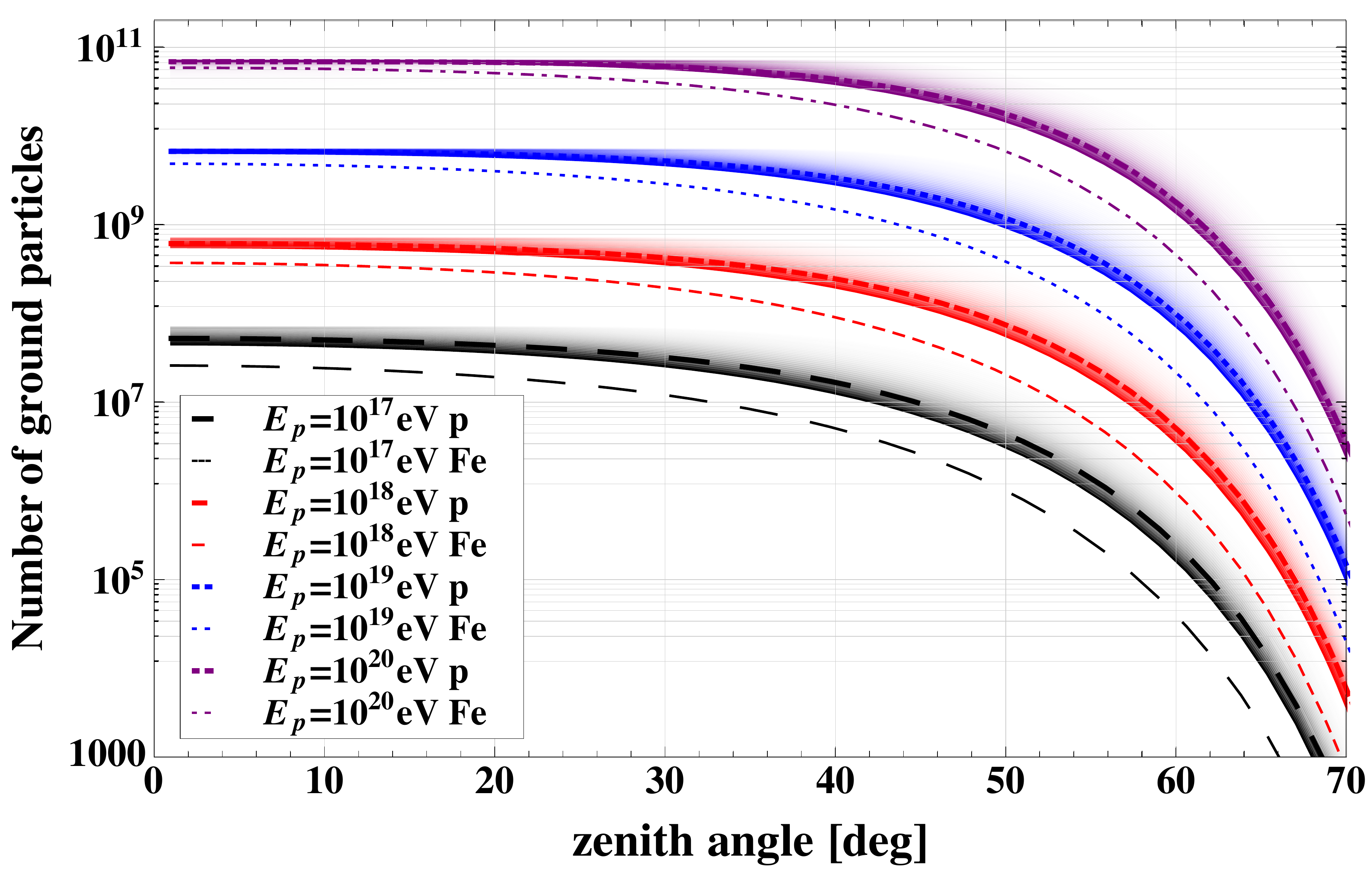}
\includegraphics[scale=0.239]{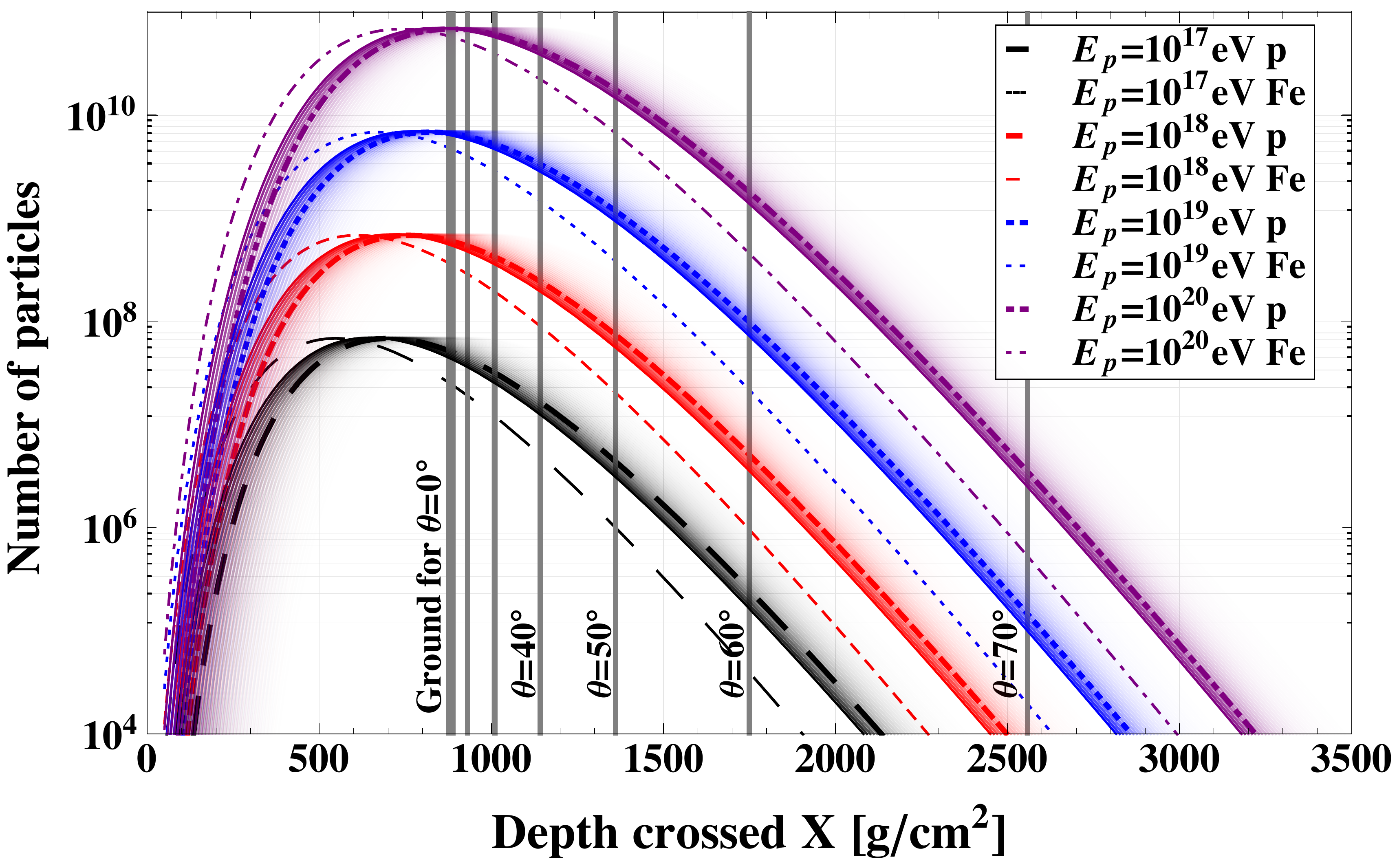}
\end{center}
\caption{\footnotesize{Top: number of particles reaching the ground $N(X_{\text{grd}})$ as a function of zenith angle, for showers initiated by proton and iron nuclei of various energies, computed for an altitude of 1400~m corresponding to the Auger site. The Greisen-Iljina-Linsley parameterization~\cite{GIL,Greisen} of the longitudinal profile is used to estimate $N(X_{\text{grd}})$. The different lines (solid, dotted, dashed...) represent the number of ground particles, obtained for showers with a value of $X_1$ set to the mean value $<X_1(E,A)>$ according to the primary  energy and mass (using the QGSJET cross section extrapolations~\cite{Ostapchenko2006143,heck-2001}). The fluctuation of $X_1$ implies a fluctuation on $N(X_{\text{grd}})$, represented by the variable color intensity. More intense means more probable.
Bottom: corresponding longitudinal profiles. The vertical gray lines represent the depth crossed by the shower until it reaches the ground, depending on the value of the zenith angle $\theta$.}}
\label{PartGround}
\end{figure}

Depending on the air shower characteristics and on the altitude of the observation site, the number of secondary particles reaching the ground $N(X_{\text{grd}})$ can vary strongly.
For a given ground altitude, $N(X_{\text{grd}})$  mainly depends on: 
\begin{itemize}
\item the energy of the primary cosmic ray inducing the cascade in the atmosphere;
\item the air shower arrival direction (zenith angle);
\item the first interaction length $X_1$ of the primary cosmic ray;
\item the nature of the primary cosmic ray.
\end{itemize}
Fig.~\ref{PartGround} (top) shows the number of particles reaching the ground as a function of arrival direction, for showers induced by proton and iron nuclei of various energies. 
The corresponding longitudinal profiles are presented in Fig.~\ref{PartGround} (bottom). The maximum number of particles $N_{\text{max}}$ in the shower is obtained, by definition, at the depth of maximum development $X_{\text{max}}$. We see that for a primary energy above $10^{18}$ eV (1~EeV) and for arrival directions $\theta\leqslant50^\circ$, $N(X_{\text{grd}})$ is not negligible with respect to $N_{\text{max}}$ ($N(X_{\text{grd}})/N_{\text{max}}\geqslant10\%$).

A large majority of secondary electrons and positrons (at least 97\%) for $X\geqslant X_{\text{max}}$, have an energy smaller than 1~GeV (see~\cite{lafebre2009}). Consequently, when the shower front hits the ground, the secondary electrons and positrons are stopped on a distance smaller than 1~m at 1~GeV in a silicon medium~\cite{Nist}, similiar to the ground at the Auger site. The radiation of electrons and positrons under ground level is strongly attenuated by the small conductivity of this medium: $\sigma^{\text{max}}_{\text{ground}} = 8.10^{-3}$~S/m at 80~MHz~\cite{soil}, implying a complete attenuation of the radiation after a few centimeters.

The very fast absorption of these particles ---~we call this phenomenon the sudden death (SD) of the shower~--- is at the origin of the emission of an electric field through various mechanisms:
\begin{itemize}
\item the sudden disappearance of the net negative charge in the shower front generated by secondary electrons in excess, and the corresponding longitudinal current;
\item the sudden disappearance of the transverse current due to systematic opposite drift of electrons and positrons in the geomagnetic field;
\item the asymetry in the ground distribution of electrons and positrons, induced by the charge separation.
\end{itemize}
This electric field is emitted by a region close to the shower core. An observer will detect the signal at a time~$\sim d/c$ after the impact on the ground, where $d$ is the distance to the shower core. The electric field emitted during the development of the shower in the atmosphere creates the PP as discussed in the introduction, which is detected before this sudden death pulse (SDP).
The time interval between the PP and the SDP depends on the geometry (the air shower arrival direction and the antenna position with respect to the ground shower core) but also on the instant of the maximum of emission during the air shower development. 

In the next section, we quantify and study in details the SDP using the simulation.

\section{III. Description of the pulse}

\subsection{Example of vertical simulated showers}

We ran simulations of proton-induced vertical showers falling at the origin of the coordinate system for energies $0.1, 0.3, 1, 3, 10, 30, 100$~EeV with the code SELFAS2 for 48 antennas located between 100~m and 800~m by steps of 50~m at the geographic east, west and north of the shower core and one antenna located at the origin. The ground level is set at 1400~m and we consider the geomagnetic field measured at the Pierre Auger Observatory.
The first interaction point ($X_1$) at a given energy is extracted from the QGSJET data as in Fig.~\ref{PartGround}, and is taken as a function of energy as $52.8, 50.8, 48.6, 46.8, 44.9, 43.2, 41.5$~g~cm$^{-2}$, respectively. The electric field received by each antenna as a function of time is calculated in the EW, north-south (NS) and vertical (V) polarizations.
The observer's origin of time $t=0$ corresponds to the time when the shower axis hits the ground at the core position~$(0,0,1400)$.

\begin{figure}[!ht]
\begin{center}
\includegraphics[width=9cm]{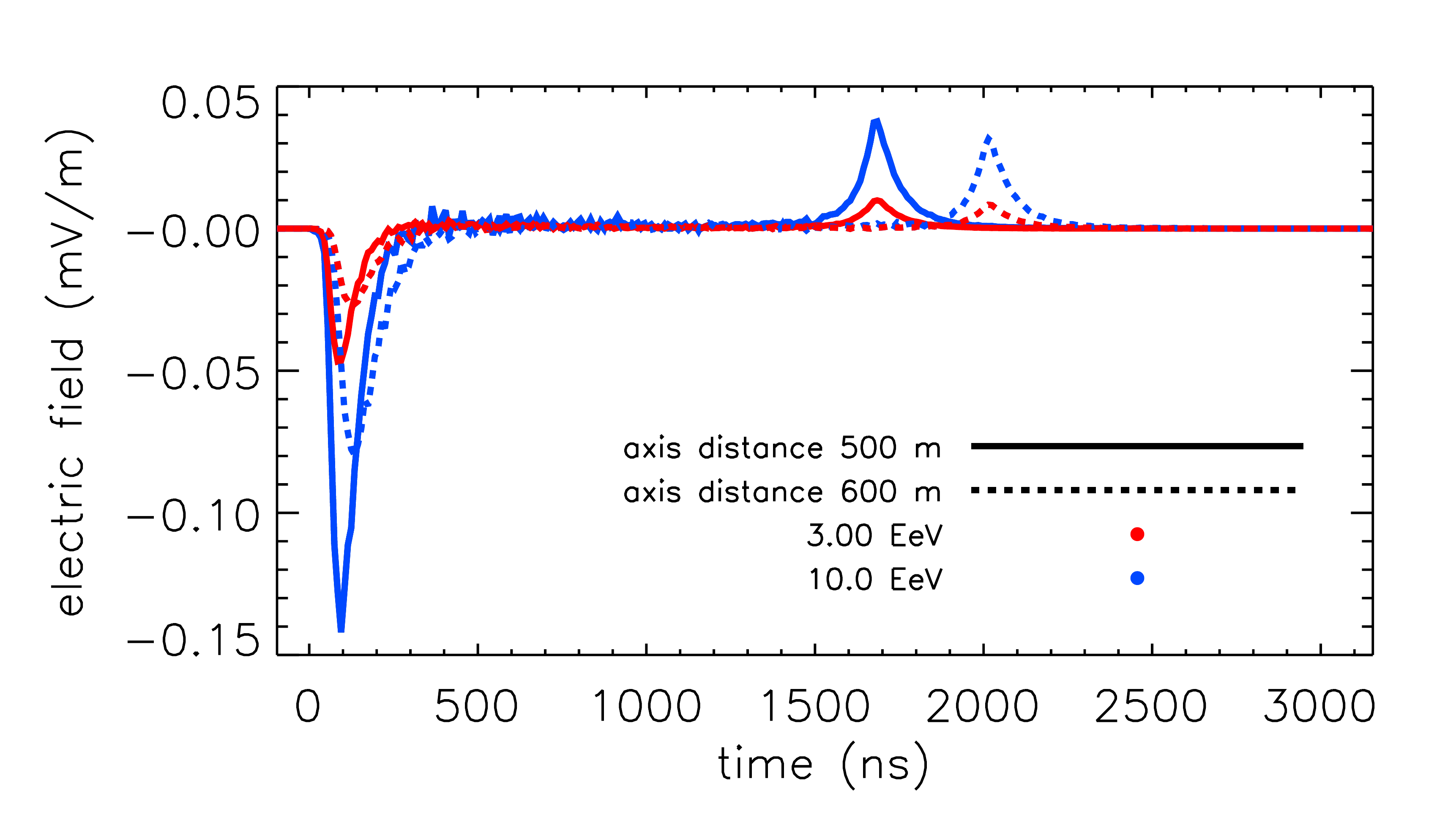}
\end{center}
\caption{\footnotesize{EW electric field vs time for two antennas located at 500~m and 600~m at the east of the shower core, for vertical showers at 3 and 10~EeV.}}
\label{pulseEWbrasE}
\end{figure}

\begin{figure}[!ht]
\begin{center}
\includegraphics[width=9cm]{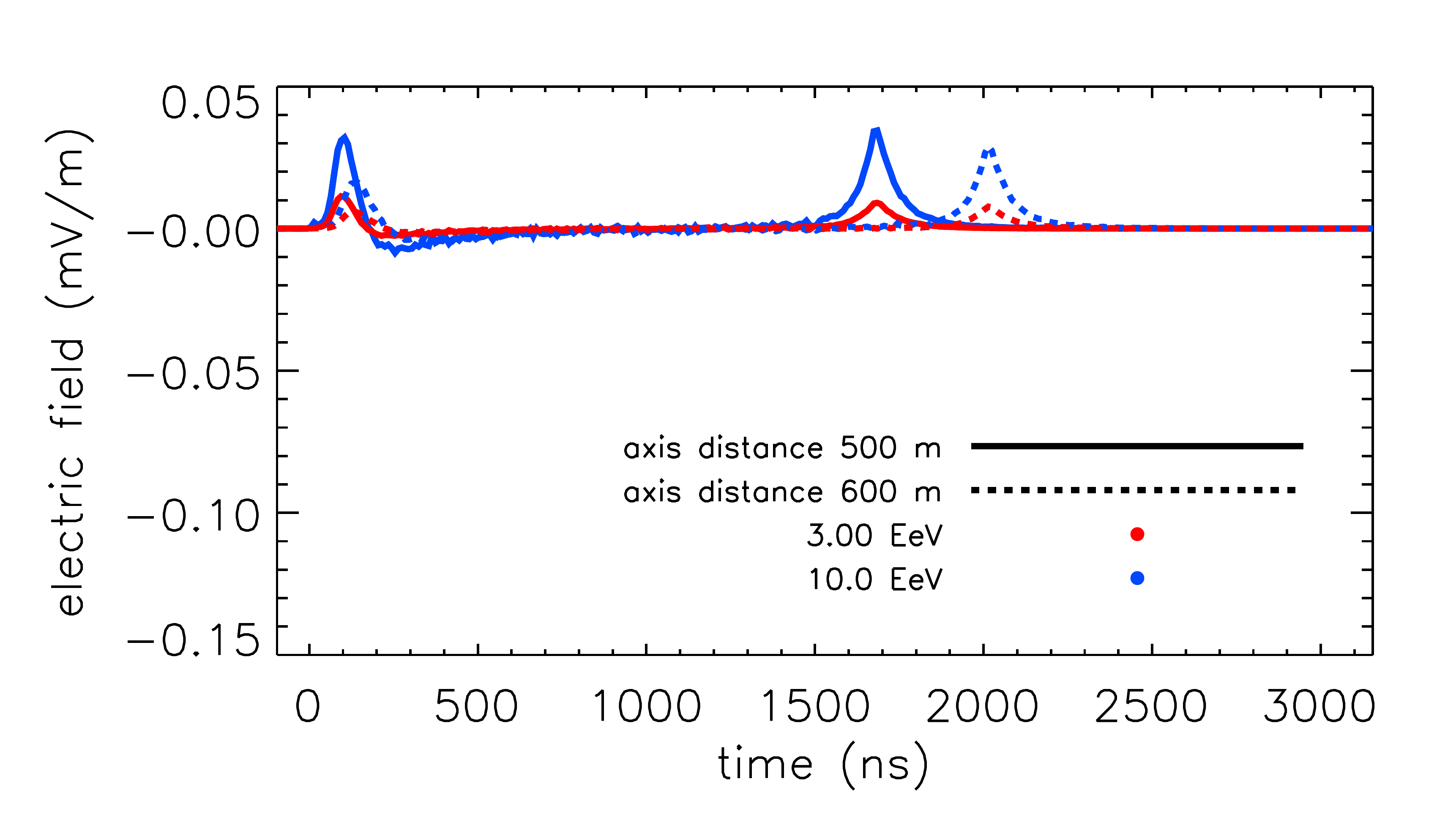}
\end{center}
\caption{\footnotesize{Same as Fig.~\ref{pulseEWbrasE} for the V polarization.}}
\label{pulseVbrasE}
\end{figure}

Figs.~\ref{pulseEWbrasE} and~\ref{pulseVbrasE} presents the EW and V electric fields, respectively, as a function of time at 500~m and 600~m at the east of the shower core.

The PP due to the various mechanisms generating the electric field in the shower during the development in the atmosphere is clearly visible at the beginning of the trace, with a maximum amplitude occuring at a time close to 100~ns in this example.
The SDP appears at roughly 1600~ns and 2000~ns for antennas located at 500~m and 600~m of the shower core, respectively. Note that in this example, no SDP is visible in the NS polarization for these antennas located at the east of the shower axis. This means that the SD electric field lies in the vertical plane containing the shower axis and the observation point, in agreement with an electric field aligned with $\mathbf{n}-\boldsymbol\beta$ (see Eq.~\ref{SumField}). For a vertical shower, an observer measuring only the electric field in the horizontal plane will reconstruct a radial polarization pattern, in agreement with Eq.~\ref{SumField} which predicts a radial polarization pattern in the horizontal plane when the shower is vertical ($\boldsymbol\beta\sim-\mathbf{z}$). We checked that the contribution to the SDP is dominated by the excess of negative charge in the shower.

\subsection{Time structure of the SDP}

From the simulation, the time interval between the time origin and the time $t^{\text{max}}_{\text{SDP}}$ when the SDP reaches its maximum value is simply given by $\delta t=d/c$, in agreement with the hypothesis that the SDP is due to the disappearance of secondary particles when hitting the ground. This result is also verified for all antennas used in this simulation.

We can infer the shape of the SDP. For that, we describe the shower front as a macroscopic spatial charge density defined by $\mathcal{Q}(x,y,z,t_{\text{ret}})$.
This function describes the thickness and lateral spread of the shower front.
Before reaching the ground, the time variation of $\mathcal{Q}$ due to the evolution of the number of particles in the shower is small compared to the charge variation induced by the sudden absorption of the particles in the ground, which is a quasi instantaneous event. The time derivative of a decreasing charge to zero explains why the SDP is monopolar, as seen in Figs~\ref{pulseEWbrasE} and~\ref{pulseVbrasE}. When the shower front hits the ground,
the observer sees the shower front disappearing at the speed of light. The duration of the complete absorption in the ground depends on the inclination of the shower, on its curvature and its thickness.
The thickness of the shower front ($\approx2$ m) is negligible compared to the size of the lateral extension (Moli\`ere radius $\approx100$ m), so that we expect that the time structure of the SDP is mainly due to the lateral spread of the shower front. The SDP time structure provides an image of the particle ground footprint.
For this reason, the shape of the pulse should be predicted quite easily by simple geometrical considerations. Two equivalent points of view can be adopted to model the shape of the SDP.

\subsubsection{Analytical description using the LDF, for a vertical shower}
\begin{figure}[!ht]
\begin{center}
\includegraphics[width=8cm]{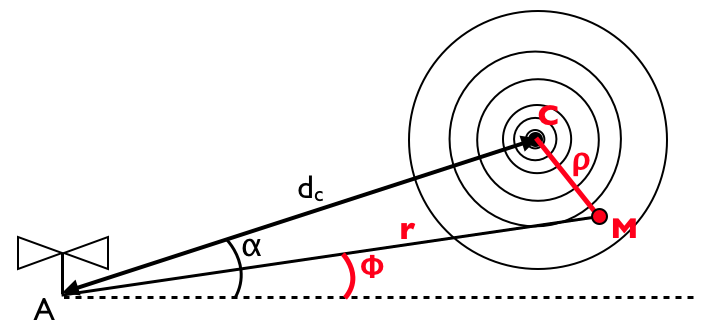}
\end{center}
\caption{\footnotesize{System of coordinate used to estimate the time structure of the SDP. $A$ is the observer position and $M$ a point on the ground where the particle density is given by the LDF $\mathcal{L}(\rho)$, $\rho$ which is the distance between $M$ and the ground shower core. }}
\label{Schema2}
\end{figure}

Let's first consider the case of a vertical shower where the ground particles density (per unit area) is well described by a symetric lateral distribution function (LDF) $\mathcal{L}(\rho)$, where $\rho$ is the distance to the shower axis.
The shower core is located at the position $(x_c,y_c)$ and the observer at the position $(x_a,y_a)$. We call $d_c$ the distance between the shower core and the observer and $\alpha$ the angle under which the observer in $A$ sees the shower core (see Fig.\ref{Schema2}). Let $M$ be a point on the ground located at a distance $r$ from $A$. At $M$, the particle density is given by the value of the LDF at this point, at a distance $\rho(r,\phi)=\sqrt{r^2+d^2_c-2rd_c\cos(\alpha-\phi)}$ from the shower core. The number of particles around $A$ in the interval $[r;r+\text{d}r]$ is given by:
\begin{equation}
\frac{\text{d}N}{\text{d}r}(r,d_c)=r\,\int_0^{2\pi}\mathcal{L}(\rho(r,\phi))\,\text{d}\phi
\end{equation}
According to Eq.~\ref{SumField}, the particles located in the interval $[r;r+\text{d}r]$ create an electric field $E_{\text{SDP}}(t_A)$ observed by $A$ proportional to $1/r$, detected at the time $t_A=r/c$:
\begin{equation}
E_{\text{SDP}}(t_A=r/c)\propto\left[\frac{1}{r}\frac{\text{d}N}{\text{d}r}(r,d_c)\right]_{\text{ret}}=\int_0^{2\pi}\mathcal{L}(\rho(r,\phi))\,\text{d}\phi\,\cdot\label{sdp2D}
\end{equation}
Using the Greisen lateral distribution function $\mathcal{L}(\rho)$ given in \cite{doi:10.1146/annurev.ns.10.120160.000431} at the ground level, we show the result obtained in Fig.~\ref{sdpmodvertical} in the case of a vertical shower. The absolute amplitude of the analytic SDP model is fitted to the simulated SDP obtained at 200~m and the same factor is used for the antenna located at 400~m from the shower core. The analytical model reproduces very well the time structure and the relative amplitude of the SDP.

\subsubsection{Description using the ground particles distribution}
\begin{figure}[!ht]
\begin{center}
\includegraphics[width=8cm]{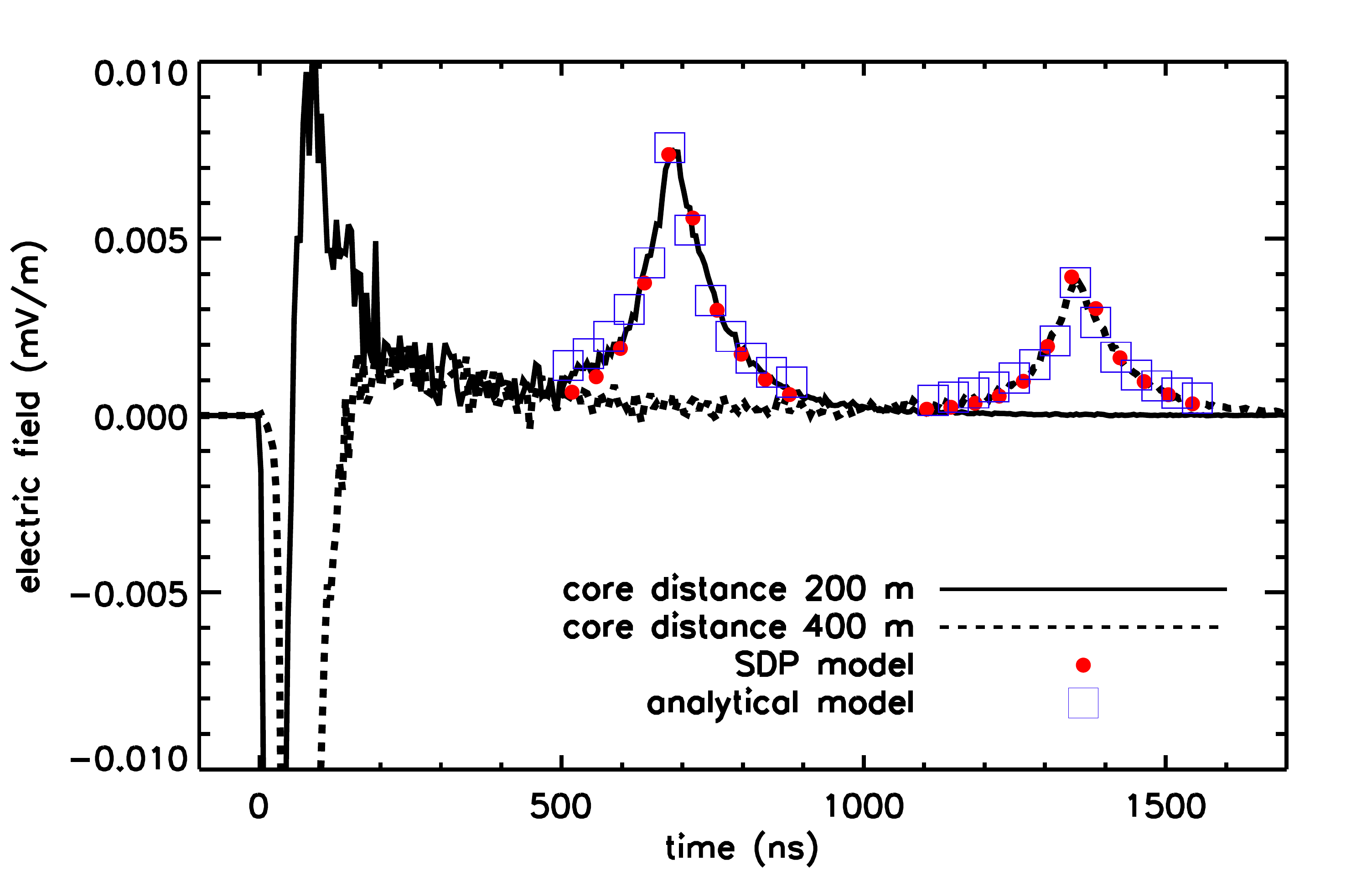}
\end{center}
\caption{\footnotesize{SDP model applied to a vertical shower at 1~EeV using two antennas in the EW polarization at 200~m and 400~m at the east of the shower core. The red dots are the predicted SDP using the model based on the ground particles distribution provided by the simulation. The blue squares correspond to the SDP predicted using the analytical Greisen formula for the LDF. In both cases, the SDP is very well reproduced.}}
\label{sdpmodvertical}
\end{figure}

\begin{figure}[!ht]
\begin{center}
\includegraphics[width=8cm]{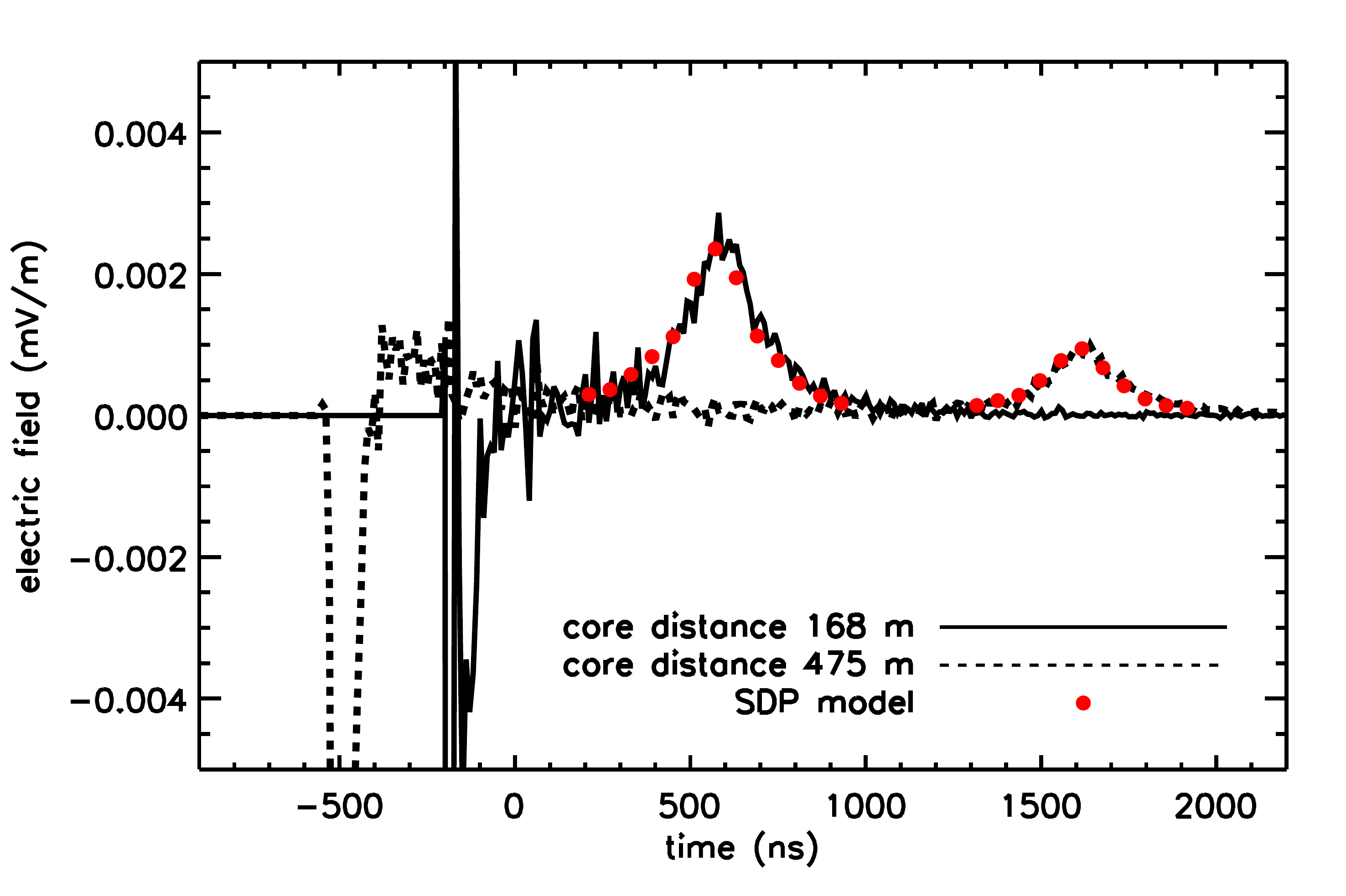}
\end{center}
\caption{\footnotesize{SDP model applied to an inclined shower at 1~EeV coming from the north-east ($\theta=30^\circ$, $\phi=45^\circ$) using two antennas in the NS polarization at 157~m and 445~m from the axis. The antennas are located at the east of the shower core, at 168~m and 475~m respectively.}}
\label{sdpmodinclined}
\end{figure}

The analytical model can be extended to inclined showers, provided we know the actual LDF, including the asymetries due to the shower inclination: the LDF depends on $\rho$ but also on the relative position around the core.
A more direct method to compute the expected SDP for inclined showers, is to use directly the information provided by the simulation. In the simulation, we record the time and ground position $(t_i,x_i,y_i,z_i=1400~\text{m})$ of each simulated particle $i$ hitting the ground. We assume that these particles have the same velocity parallel to the shower axis $\boldsymbol\beta$. According to Eq.~\ref{SumField}, a given ground particle $i$ located at the point $M$ of coordinates $(t_i,x_i,y_i,z_i=1400~\text{m})$ creates an electric field at the position $A$ proportional to $(\mathbf{n}-\boldsymbol\beta)/R_i$ where $R_i=||\boldsymbol{MA}||$ and $\mathbf{n}=\boldsymbol{MA}/R_i$. This signal is detected by the observer at the time $t_i^A=t_i+R_i/c$.
For this observer, the total signal detected in the time interval $[t_i^A;t_i^A+\delta t[$ is the summation of all contributions of the ground particles verifying $t_i^A\leqslant t_i+R_i/c< t_i^A+\delta t$. This simple modelization of the SDP can be compared to the result of the simulation. In Fig.~\ref{sdpmodvertical} we show the simulated electric field for two antennas located at 200~m and 400~m at the east of the shower axis for a vertical shower of 1~EeV. In Fig.\ref{sdpmodinclined}, we show the same analysis for an inclined shower coming from the north-east ($\theta=30^\circ$, $\phi=45^\circ$) in the NS polarization for two antennas at the north of the shower core, located at 168~m and 475~m, respectively (axis distances 157~m and 445~m). We see that the agreement is very good. For a given simulated shower (either vertical or inclined as shown in Figs~\ref{sdpmodvertical} and \ref{sdpmodinclined}), the absolute amplitude of the SDP model is fitted to the simulated SDP and the same factor is used for all antennas corresponding to this shower. This means that the model reproduces correctly both the time structure and the relative amplitude of this signal.

\begin{figure}[!ht]
\begin{center}
\includegraphics[width=9cm]{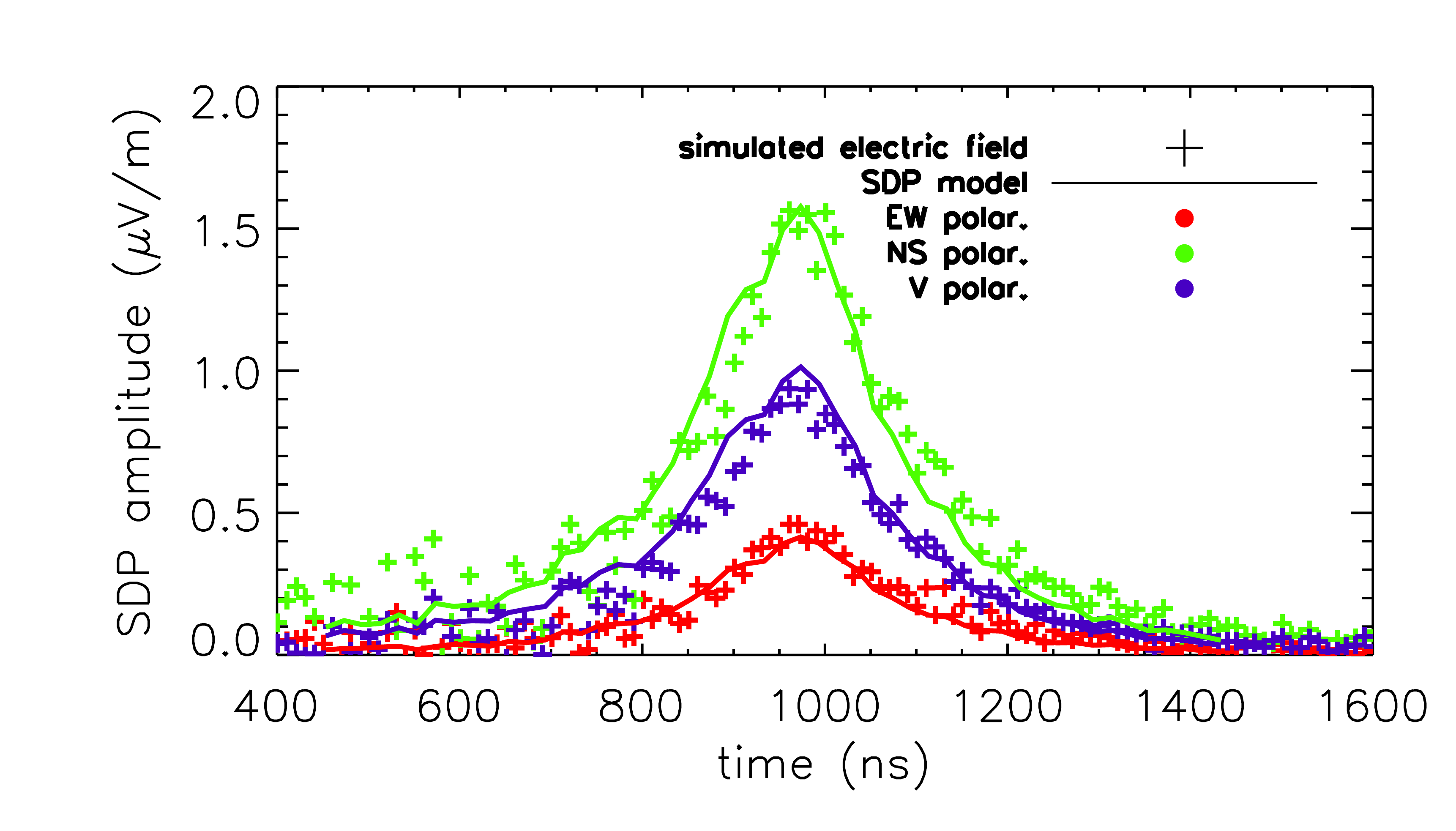}
\end{center}
\caption{\footnotesize{SDP model applied to the three polarizations of a single antenna located at 268~m at the east of the shower core. The same normalization factor is used for the three polarizations. The shower is coming from $\theta=30^\circ,~\phi=45^\circ$, the primary energy is 1~EeV.}}
\label{sdpmod3pol}
\end{figure}

For a given antenna, Fig.~\ref{sdpmod3pol} shows that the model predicts correctly the SDP in the three polarizations. The data presented here correspond to the simulated electric field computed at 268~m at the east of the shower core, for an inclined shower $\theta=30^\circ$ coming from the north-east, at 1~EeV. The SDP model is computed for the three polarizations and the same normalization factor is used, this confirms that the SDP pulse is well described by an electric field along $\mathbf{n}-\boldsymbol\beta$.

\subsection{SDP amplitude characterization}
In this section we characterize the SDP amplitude ${\cal A}$ as a function of energy and core distance. We estimate the amplitude by fitting the SDP with the function:
\begin{equation}
\text{SDP}(t) = {\cal A}\exp\left(-\frac{|t-t^{\text{max}}_{\text{SDP}}|}{\tau_\pm}\right)
\end{equation}
where $t^{\text{max}}_{\text{SDP}}$ is the time when the SDP reaches its maximum value and $\tau_\pm$ is a time constant which can take different values on the two intervals $t\leqslant t^{\text{max}}_{\text{SDP}}$ and $t>t^{\text{max}}_{\text{SDP}}$.

\subsubsection{Vertical case}

\begin{figure}[!ht]
\begin{center}
\includegraphics[width=9cm]{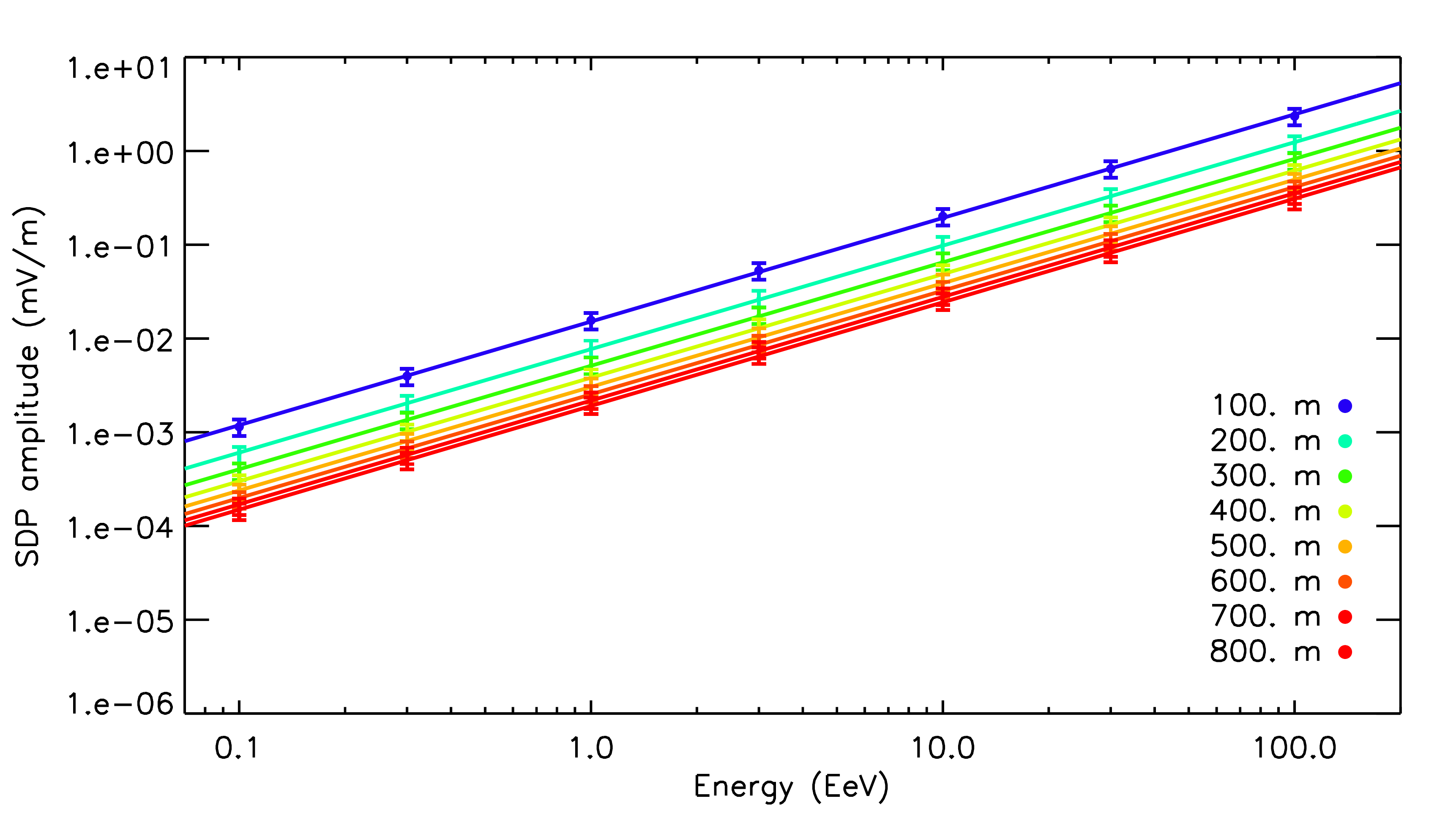}
\end{center}
\caption{\footnotesize{SDP amplitude in the EW polarization as a function of energy for different core distances. The amplitude varies almost linearly with the energy.}}
\label{ampvsenergy}
\end{figure}

Fig.~\ref{ampvsenergy} presents the dependence of ${\cal A}$ with the primary energy, for observers located at different core distances, at the east of the shower axis. The depence is almost linear.

\begin{figure}[!ht]
\begin{center}
\includegraphics[width=9cm]{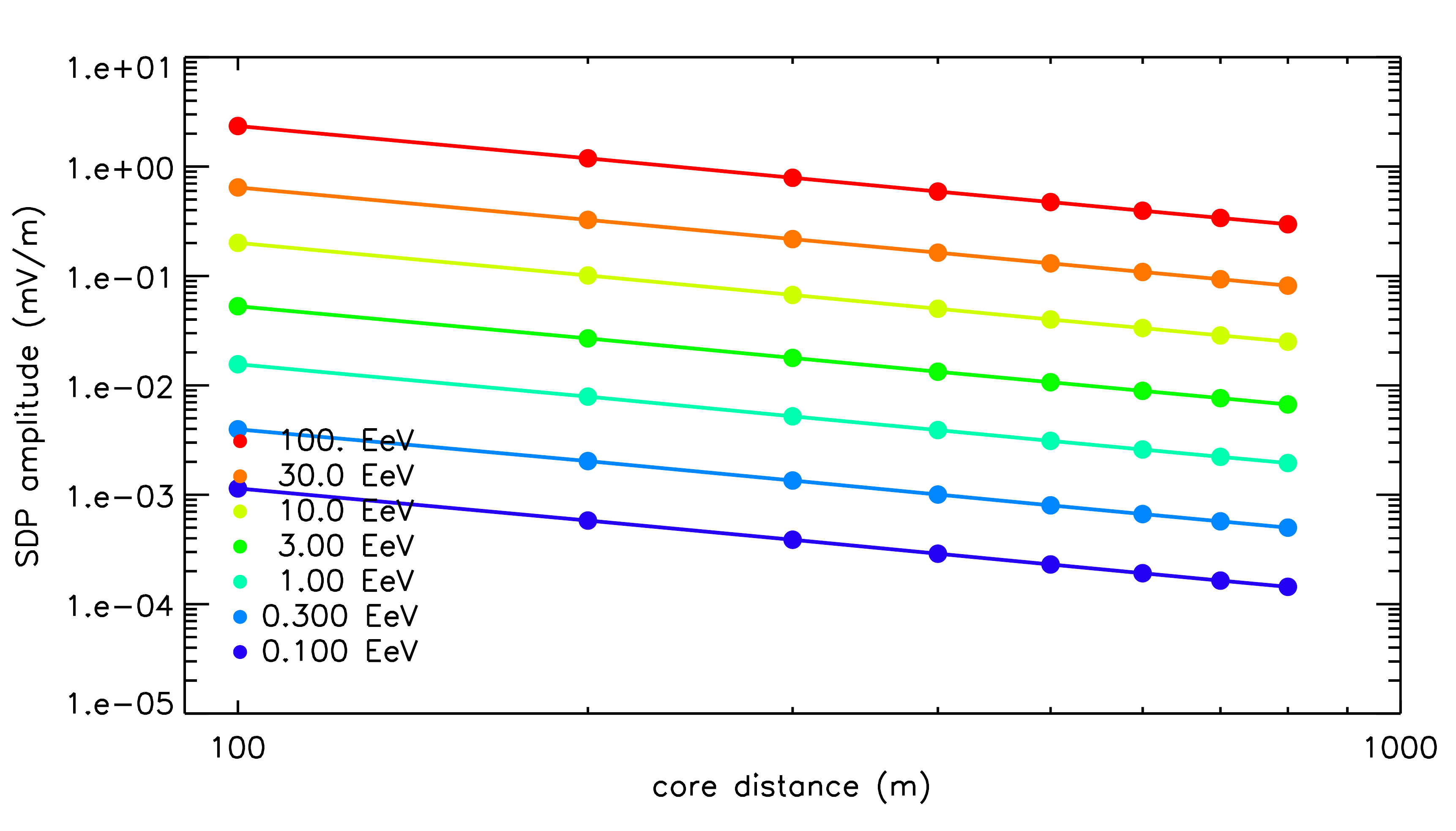}
\end{center}
\caption{\footnotesize{SDP amplitude in the EW polarization as a function of core distance for different energies. The amplitude scales as $1/d$.}}
\label{ampvsdistance}
\end{figure}

Fig.~\ref{ampvsdistance} presents the dependence of ${\cal A}$ with core distance, at different energies. We see that ${\cal A}$ scales as the inverse of the core distance, proving the radiative nature of the SD emission as inferred in the previous section.
We can finally propose a complete parameterization of the SDP amplitude ${\cal A}$ as a function of energy and core distance:
\begin{equation}
{\cal A}^{\text{pol}}_{\text{SDP}}(E,d)={\cal A}^{\text{pol}}_{\text{norm}}\left(\frac{100~\text{m}}{d}\right)\left(\frac{E}{1~\text{EeV}}\right)^{1.1}~\mu\text{V}\,\text{m}^{-1}\label{eqscale}
\end{equation}
the factor ${\cal A}^{\text{pol}}_{\text{norm}}$ being the normalized amplitude in a given polarisation ("pol" being EW, NS or V) at 1~EeV and a reference core distance of 100~m.
For instance, in the EW polarization for observers at the east of the shower axis, ${\cal A}^{\text{EW}}_{\text{norm}}=15.6~\mu\text{V}\,\text{m}^{-1}$. We find the same order of amplitude
for antennas at the west of the shower core in the EW polarization and at the north in the NS polarization.
Note that the SDP amplitude is relatively higher in the V polarization than in the horizontal polarizations when compared to the PP amplitude.
At fixed energy, the SDP amplitude decreases less slowly than the PP amplitude so that the ratio of the amplitudes SDP/PP can even be greater than~1 (but at energies above 20~EeV for horizontal polarizations).

\subsubsection{Inclined case}
\begin{figure}[!ht]
\begin{center}
\includegraphics[width=9cm]{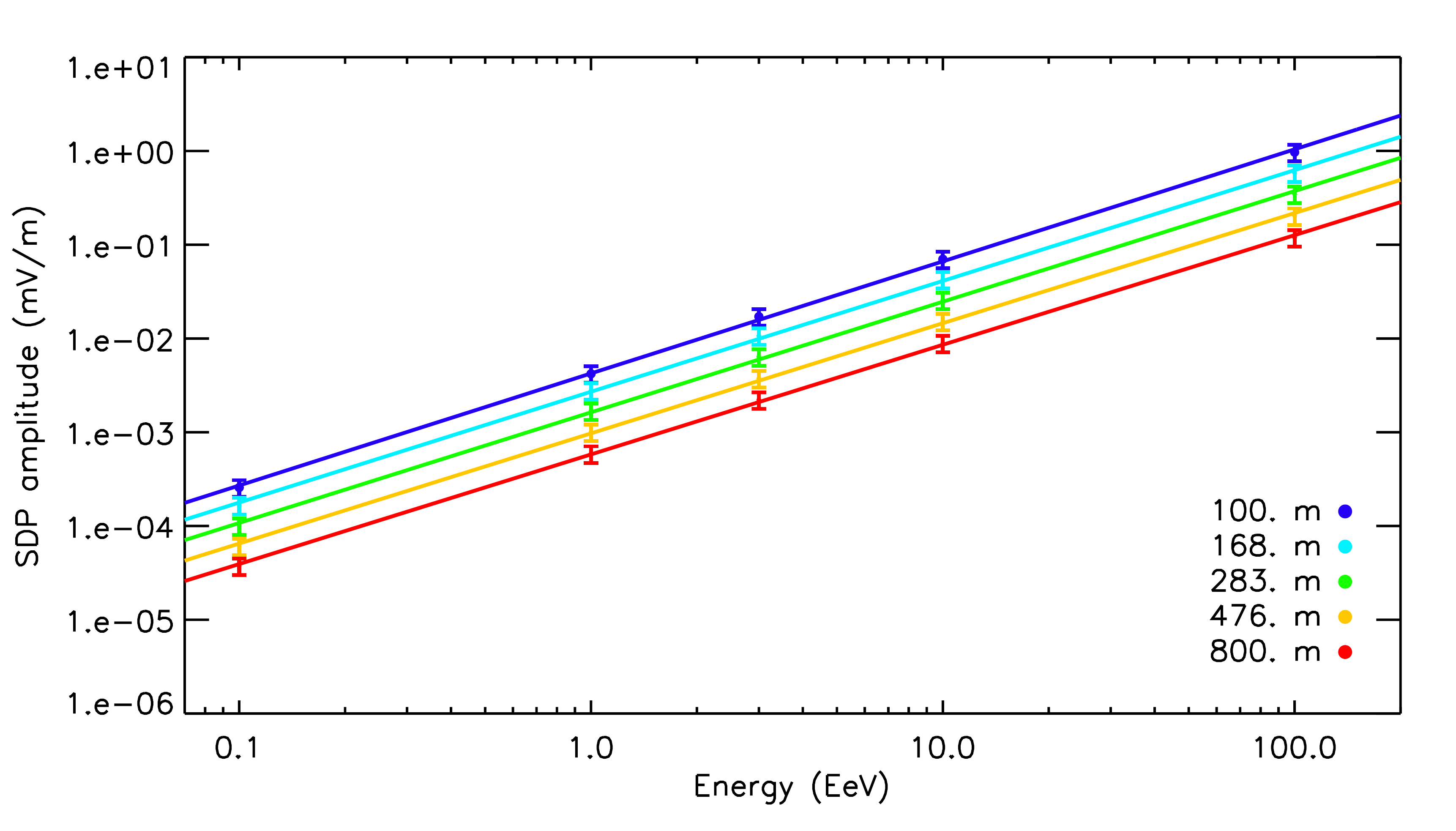}
\end{center}
\caption{\footnotesize{SDP amplitude in the NS polarization as a function of energy for different core distances. The antennas are located at the north of the shower core. The SDP amplitude varies almost linearly with the energy. This case corresponds to an inclined shower of energy 1~EeV, with $\theta=30^\circ$ and $\phi=45^\circ$.}}
\label{calibinclined}
\end{figure}

At fixed energy, first interaction point and primary cosmic ray, the number of particles reaching the ground is maximum when the shower is vertical (see Fig.~\ref{PartGround}). Therefore, we expect a smaller amplitude for inclined showers. The amplitude scales in the same way as vertical showers (see Eq.~\ref{eqscale}), in both energy and core distance; the reference amplitude at 1~EeV for a core distance of 100~m is of the order of 5~$\mu$V/m for a zenith angle of 30$^\circ$. Fig.~\ref{calibinclined} presents the variation of the SDP amplitude $\cal A$ for an inclined shower with $\theta=30^\circ,~\phi=45^\circ$, as a function of energy for different core distances.

\subsection{Frequency domain}

\begin{figure}[!ht]
\begin{center}
\includegraphics[width=9cm]{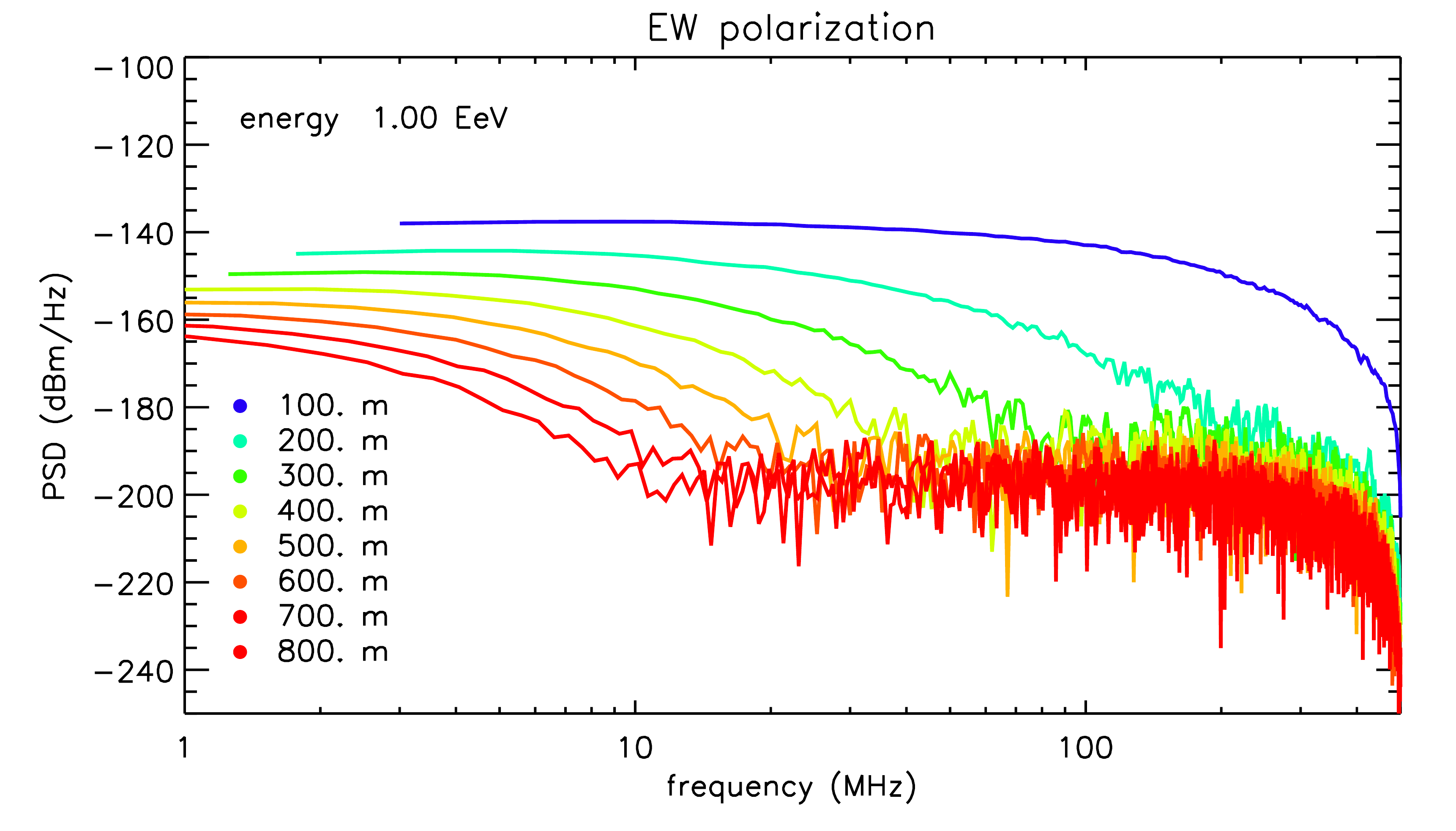}
\end{center}
\caption{\footnotesize{PSD of the PP in the EW polarization for different core distances located at the east of the shower axis. This PSD corresponds to a vertical shower of energy 1~EeV.}}
\label{psdpp}
\end{figure}

\begin{figure}[!ht]
\begin{center}
\includegraphics[width=9cm]{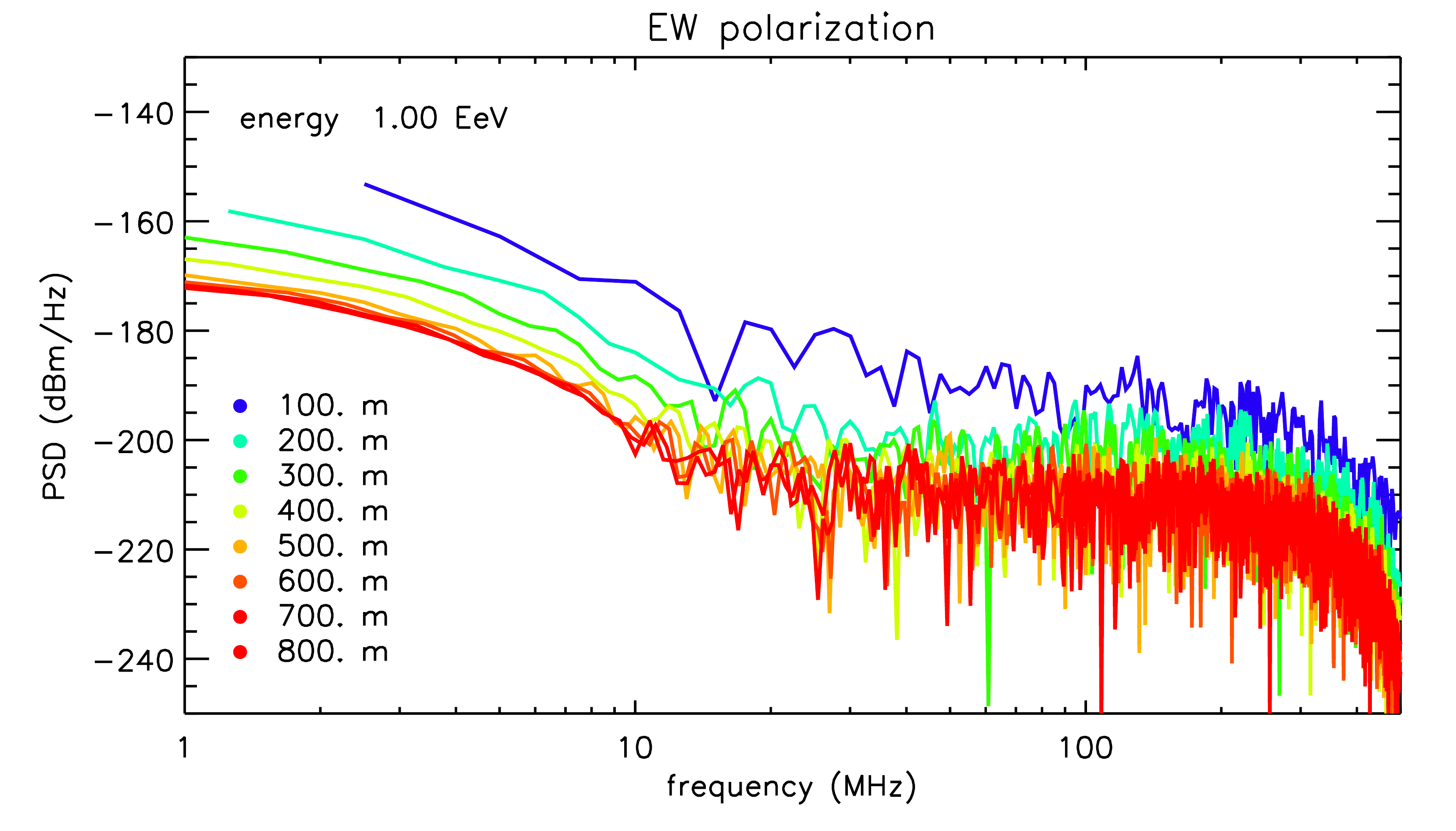}
\end{center}
\caption{\footnotesize{PSD of the SDP in the EW polarization for different core distances located at the east of the shower axis. This PSD corresponds to a vertical shower of energy 1~EeV.}}
\label{sdpsp}
\end{figure}

Contrarily to the time structure of the PP, that of the SDP does not depend on the primary energy nor on the distance to the shower core as can be seen in Figs.~\ref{pulseEWbrasE} and~\ref{pulseVbrasE}. Only the amplitude is affected by these parameters. Therefore, we expect the shape of the power spectral density (PSD) to be similar at all core distances and all energies as a function of frequency, the only difference being the normalization of the PSD. Fig.~\ref{psdpp} presents the PSD of the PP for a primary energy of 1~EeV and core distances varying from 100~m to 800~m. We observe, as already predicted by several simulation codes, that the PP PSD is coherent up to a certain frequency depending on the axis distance of the observer.
This is not the case for the SDP since we observe that its PSD is coherent up to a frequency of $\sim20$~MHz whatever the core distance is, as can be seen in Fig.~\ref{sdpsp}.

\section{Discussion and conclusion}

\begin{figure}[!ht]
\begin{center}
\includegraphics[width=9cm]{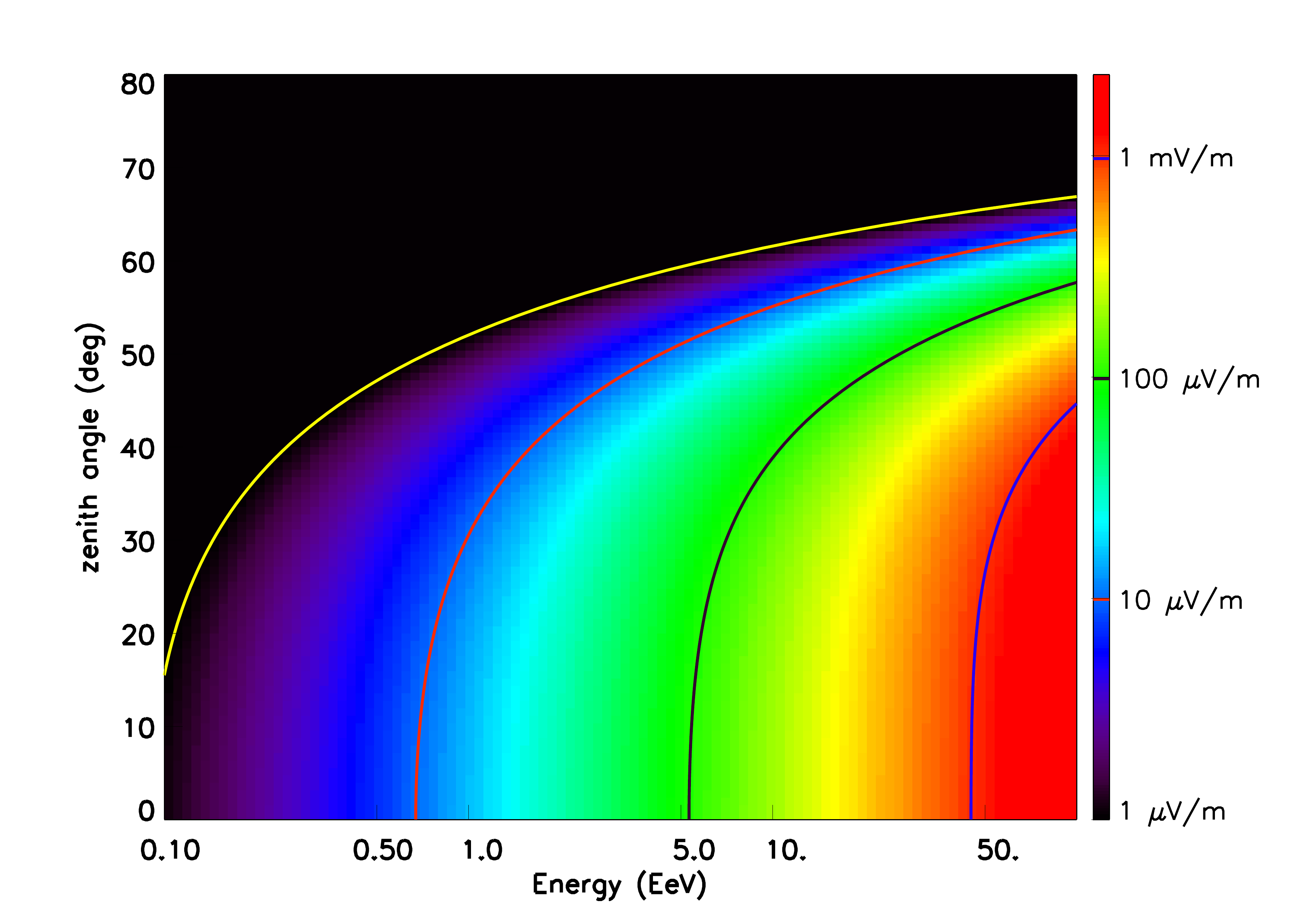}
\end{center}
\caption{\footnotesize{Expected amplitude of the SDP at 100~m at the east of the shower core, in the EW polarization, as a function of primary energy and zenith angle, computed for the site of the Pierre Auger Observatory.}}
\label{ampEt}
\end{figure}

We have characterized the electric field produced by the secondary $e^\pm$ when they reach the ground. The amplitude of this signal in the horizontal polarization, is of the order of $15~\mu$V/m at 100~m of the shower core for a vertical shower,
and a primary energy of 1~EeV. The amplitude scales quasi-linearly with the energy and decreases as $1/d$, where $d$ is the distance between the shower core and the observer. The SDP amplitude depends on the distance to the shower core, contrarily to the PP due to the secondary $e^\pm$ during the shower development, which strongly depends on the distance to the shower axis. The polarization is oriented along the vector $\mathbf{n}-\boldsymbol\beta$, where $\mathbf{n}$ is the normalized vector between the shower core and the observer position and $\boldsymbol\beta$ can be taken as the shower axis direction. The signal exists if the number of $e^\pm$ in the shower at the ground level is high enough: for a given observation site, the condition to observe the SDP depends on the zenith angle and primary energy of the shower. Fig.~\ref{ampEt} presents the expected SDP amplitude at 100~m at the east of the shower core, in the EW polarization, as a function of zenith angle and energy. We assumed proton-initiated showers with the first interaction point $X_1$ set to the average value extracted from QGSJET data as a function of primary energy.
At the Pierre Auger Observatory site for instance, the ground altitude of 1400~m implies that a total of $\sim 10^{9}$ $e^\pm$ can reach the ground for a vertical shower of 1~EeV. The SDP power spectral density is important at relatively low frequencies, below 20~MHz.
If we are able to detect a signal at these frequencies, then it would be quite simple to trigger on actual cosmic rays because any trace having two pulses separated by some $\mu$s could be an excellent candidate. This would be a very specific signature of cosmic rays and would help a lot in discarding background events. The SDP is expected to arrive at the observer location at a delayed time $d/c$ with respect to the core time. The core is therefore located on a circle centered on the observer position and of radius $d/c$. Using the information from several detectors, the shower core is at the intersection of the corresponding circles. The SDP amplitude is proportionnal to the total number of secondary $e^\pm$ and also reflects their complete ground distribution (see Eq.~\ref{sdp2D}), contrarily to particle detectors that sample at specific locations this ground distribution.
Experimentally, one should be interested in recording the electric field over a duration greater than some $\mu$s to be able to observe the SDP (up to $\sim3~\mu$s after the impact time at 1~km from the shower core). The antennas used should be also sensitive to horizontal directions.

\begin{figure}[!ht]
\begin{center}
\includegraphics[width=9cm]{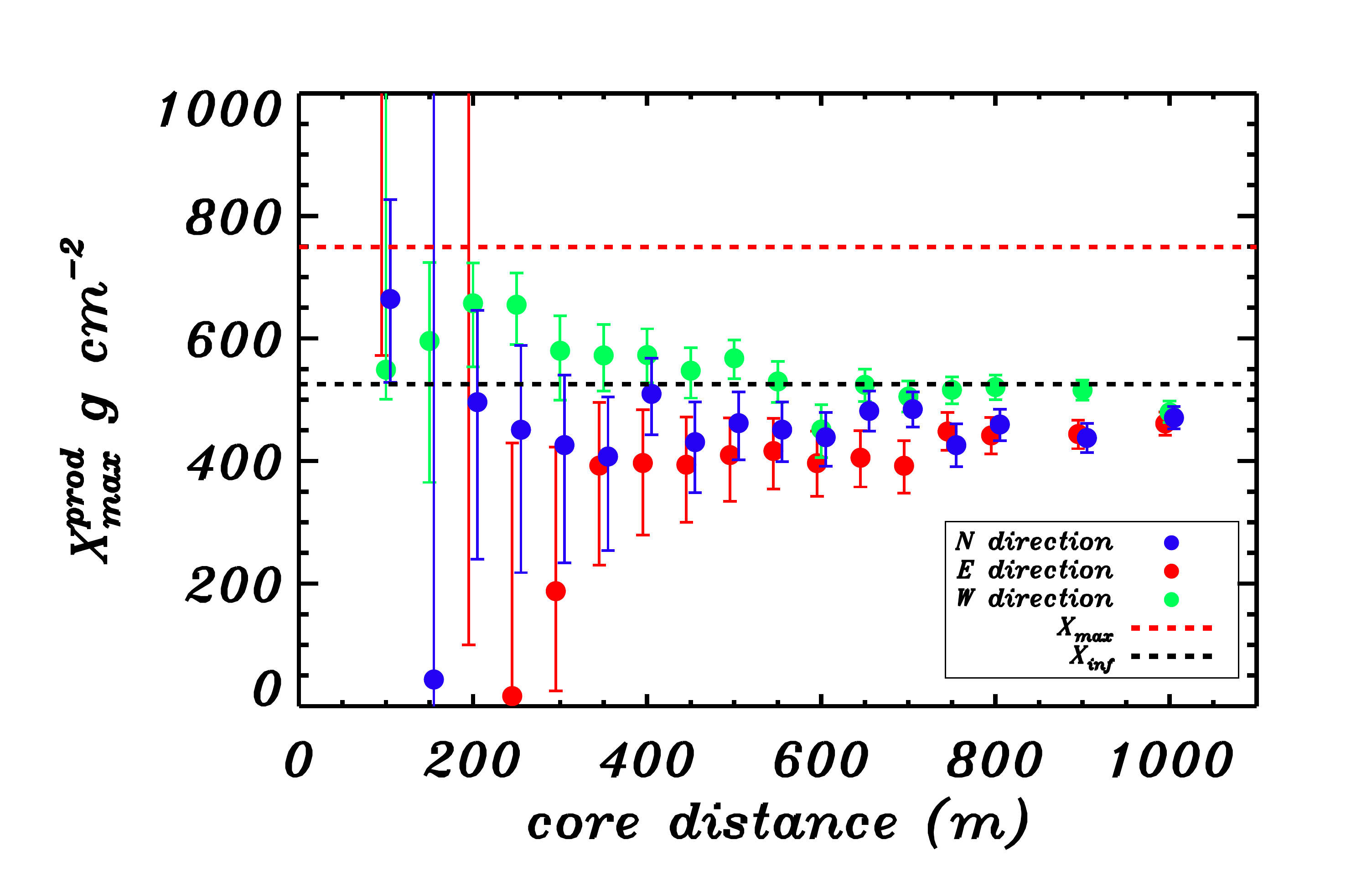}
\end{center}
\caption{\footnotesize{Estimation of $X_{\text{max}}^{\text{prod}}$ as a function of the distance to the shower core, for antennas located at the east, west and north of the core. The red dashed line indicates the $X_{\text{max}}=749~\text{g cm}^{-2}$ for this vertical simulated shower at 1~EeV. The black dashed line, indicate the depth corresponding to the first inflexion point of the longitudinal profile, $X_{\text{inf}}=525~\text{g cm}^{-2}$.}}
\label{xmaxprod}
\end{figure}

Because the SDP is generated by the end of the shower, its detection provides an absolute timing of the shower in the trace recorded by an antenna.
For antennas not too close from the shower axis (because the effect of the air refractive index is important close to the shower axis), we expect a one-to-one correspondence~\cite{PhysRevLett.107.061101} between the time in the antenna trace and the position of the source in the shower. This bijective relation can be established by simple geometrical considerations, if we assume that the shower front is point-like and moves at the speed of light~$c$.
Let $t_i$ be the time of the bin $i$ in the time trace and $t_{\text{SDP}}^{\text{max}}$ the time of the maximum amplitude of the SDP. The time $t_{\text{SDP}}^{\text{max}}$ corresponds to the end of the shower development, so that $t_i$ corresponds to the instant when the shower front was at a distance $\ell(t_i)$ from the shower core, along the shower axis, given by:
\begin{equation}
\ell(t_i)=\frac{c^2\Delta t_i^2-2c\,\Delta t_i\,||\boldsymbol{AC}|| }{2(c\Delta t_i-||\boldsymbol{AC}||+\boldsymbol{AC}\cdot \boldsymbol{u})},\label{showergeo}
\end{equation}
where $\Delta t_i=t_{\text{SDP}}^{\text{max}}-t_i$, $\boldsymbol{AC}$ is the vector from the antenna position $A$ to the shower core $C$ and $\boldsymbol{u}$ is the unit vector aligned with the shower axis $(\sin\theta\cos\phi,\sin\theta\sin\phi,\cos\theta)$.
If we consider the particular time bin corresponding to the maximum amplitude of the PP, $t_i=t_{\text{PP}}^{\text{max}}$, we can estimate the position $\ell_{\text{max}}^{\text{prod}}=\ell(t_{\text{PP}}^{\text{max}})$ along the shower axis corresponding to the maximum production of electric field, for this antenna. We finally compute the corresponding altitude above sea level $z_{\text{max}}^{\text{prod}}$ taking into account the Earth's curvature, and we finally obtain the associated depth $X_{\text{max}}^{\text{prod}}$, assuming the US standard model for the atmosphere~\cite{Sciutto:1999jh}. Fig.~\ref{xmaxprod} shows the estimation of $X_{\text{max}}^{\text{prod}}$ as a function of the distance to the shower core, for antennas located at the east, west and north of the core, for a vertical shower at 1~EeV. The error bars were computed by Monte-Carlo assuming an error on $\Delta t=t_{\text{SDP}}^{\text{max}}-t_{\text{PP}}^{\text{max}}$ varying linearly from 10~ns at 100~m of the shower core up to 20~ns at 1000~m of the shower core.
In this figure, we see two important characteristics: a convergence of the $X_{\text{max}}^{\text{prod}}$ estimation with increasing core distances and a systematic offset between the tree curves corresponding to antennas located at the north (blue), east (red) and west (green) of the shower core. The convergence can be explained by the fact that for an antenna close to the shower axis, the air refractive index generates a time dispersion of the signal which destroys the one-to-one correspondence between the time in the antenna trace and the corresponding source position in the shower ($t_{\text{ret}}$ is a non-monotonic function of $t$, see \cite{PhysRevLett.107.061101}). In that case, the relation given in Eq.~\ref{showergeo} used to estimate $X_{\text{max}}^{\text{prod}}$ cannot be applied and the estimation of the $X_{\text{max}}^{\text{prod}}$ is not valid. When the axis distance increases, the relation between $t_{\text{ret}}$ and $t$ becomes monotonic and Eq.~\ref{showergeo} can be used to estimate $X_{\text{max}}^{\text{prod}}$. The systematic offset between the three directions north, east and west is due the systematic difference in the position in the time trace of the PP. In this example of a vertical shower at 1~EeV with the geomagnetic field of the Pierre Auger Observatory site, the PP is systematically delayed for antennas located at the west of the shower axis with respect to antennas located at the north and east of the shower axis: $t_{\text{PP,west}}^{\text{max}}>t_{\text{PP,east,north}}^{\text{max}}$ at a fixed core distance.
The reason for that is the following: the two main mechanisms at the origin of the PP (respectively charge excess and transverse current variation during the shower development) produce two electric fields $\boldsymbol{E_n}$ (aligned along $\boldsymbol{n}$) and $\boldsymbol{E_{\beta^{e^\pm}_\perp}}$ (aligned along $\boldsymbol{\beta^{e^\pm}_\perp}$) that interfer constructively or destructively following the antenna position with respect to the shower axis and to the geomagnetic field. In the configuration adopted in this paper, the interference is destructive at the west of the shower core, constructive at the east of the shower core, implying a systematic relative time shift between the time of maximum of the PP amplitudes between antennas located at the east and west of the shower core.
This explains the systematic shift to higher values of $X_{\text{max}}^{\text{prod}}$ for antennas at the west of the shower core.
Fig.~\ref{xmaxprod} clearly shows that the $X_{\text{max}}^{\text{prod}}$ is not in agreement with the depth $X_{\text{max}}$ where the number of particles in the shower is maximum. The maximum of radio emission is close to the depth $X_{\text{inf}}$ at which the production rate of particles in the shower per g~cm$^{-2}$ is maximum (it corresponds to the first inflexion point of the longitudinal profile, $\left[\text{d} N/\text{d} X\right]_{X_{\text{inf}}}=0$).

\bibliography{BiblioBR}

\end{document}